\documentclass[12pt,preprint]{aastex}
\usepackage{natbib,graphics,graphicx,color}
\hfuzz=\maxdimen
\newcommand{\mum} {\micron}

\newcommand{\etal}{\hbox{et~al.\,}}
\newcommand{\psim}{\lower.5ex\hbox{$\; \buildrel \propto \over \sim \;$}}

\newcommand{\js}{}
\newcommand{\eg}{{\js e.g.\/}}

\newcommand{\Msun}{\mbox{M$_\odot$}}
\newcommand{\Lsun}{\mbox{L$_\odot$}}
\newcommand{\LIR}{\mbox{$L_{\rm IR}$}}
\newcommand{\FMI}{\mbox{$F({\rm 24~\micron})$}}
\newcommand{\0}{\phantom{0}}  

\newcommand{\AKARI}{{\it AKARI}}
\newcommand{\TMASS}{{\it 2MASS}}
\newcommand{\s}{{\it Spitzer}}
\newcommand{\h}{{\it Herschel}}

\slugcomment{\today} 
\shorttitle{Spectroscopy of $z\sim2$ Star-Forming Galaxies}
\shortauthors{Fang \etal}
\begin{document}
\title{Selection and Mid-infrared Spectroscopy
of Ultraluminous Star-Forming Galaxies at $z\sim2$}  

\author{
Guanwen Fang\altaffilmark{1,2,3}, Jia-Sheng Huang\altaffilmark{4,5,6}, 
S. P. Willner\altaffilmark{6}, Xu Kong\altaffilmark{1,7},\\
Tao Wang\altaffilmark{3,8}, Yang Chen\altaffilmark{1,9}, 
and Xuanbin Lin\altaffilmark{1,7}
}
\affil{}
\altaffiltext{1}{Center for Astrophysics, University of Science and
    Technology of China, Anhui, 230026, China}
\altaffiltext{2}{Institute for Astronomy and History of Science and Technology, 
Dali University, Yunnan, 671003, China}
\altaffiltext{3}{Key Laboratory of Modern Astronomy and Astrophysics 
(Nanjing University), Ministry of Education, Nanjing 210093, China}
\altaffiltext{4}{National Astronomical Observatories of China, Chinese Academy of Sciences, Beijing 100012, China}
\altaffiltext{5}{China-Chile Joint Center for Astronomy, Chinese Academy of Sciences, Camino El Observatorio, \#1515, Las Condes, Santiago, Chile}
\altaffiltext{6}{Harvard-Smithsonian Center for Astrophysics, 60 
Garden Street, Cambridge, MA02138, USA}
\altaffiltext{7}{Key Laboratory for Research in Galaxies and 
Cosmology, USTC, Chinese Academy of Sciences, China}
\altaffiltext{8}{School of Astronomy \& Space Science, Nanjing University, 
Nanjing 210093, China}
\altaffiltext{9}{Astrophysics Sector, SISSA, Via Bonomea 265, I-34136 Trieste, Italy}
\email{jhuang@cfa.harvard.edu, xkong@ustc.edu.cn}

%
%

\begin{abstract}
  Starting from a sample of 24~\micron\ sources in the Extended Groth
  Strip, we use 3.6 to 8~\micron\ color criteria to select
  ultraluminous infrared galaxies (ULIRGs) at $z\sim2$. Spectroscopy
  from 20--38~\micron\ of 14 objects verifies their nature and gives
  their redshifts. Multi-wavelength data for these objects imply
  stellar masses ${>}10^{11}$~\Msun\ and star formation rates
  $\ge$410~\Msun~yr$^{-1}$. Four objects of this sample observed at
  1.6~\micron\ (rest-frame visible) with {\it HST}/WFC3 show diverse
  morphologies, suggesting that multiple  formation processes create
  ULIRGs.  Four of the 14 objects show signs of active galactic
  nuclei, but the luminosity appears to be dominated by star
  formation in all cases.

\end{abstract}


\section{Introduction}\label{sec:intro}

Understanding when and how the most massive galaxies in the universe
formed is one of the chief problems in cosmology. A critical redshift
era for understanding galaxy formation is
$z\sim2$  or ``cosmic high 
noon'' \citep{grogin2011}.  In this era,
the cosmic star formation rate density begins to decline
from a flat plateau at higher redshifts, the
morphological type mix of field galaxies changes, and
the number density of quasi-stellar objects (QSOs)  peaks. About
50--70\% of the stellar mass assembly of galaxies took place in
the redshift range of $1<z<3$ \citep{dickinson2003, fontana2003,
  steidel2004, kong2006, richards2006, arnouts2007, pozzetti2007,
  noeske2007}.

Despite the importance of studying the $z\sim2$ era, practical identification
of galaxies at this redshift using only visible observations is difficult.
Visible spectroscopy in particular is greatly hampered because the familiar
spectral features shift out of the visible band while the Lyman alpha
line has not moved in yet. Nevertheless, there have been some
pioneering studies using visible-light
selection of galaxies at $z\sim2$, namely BM/BX sources 
\citep{steidel2004}, which are identified using their
rest-ultraviolet (UV) absorption lines.  
Recent near-infrared (NIR) studies have found many more
galaxies at $z\sim2$ with much higher stellar masses and more intensive
star formation \citep{daddi2004, labbe2005, papovich2006,
  dunne2009}, and infrared (IR) imaging and spectroscopy are essential
for studying the whole galaxy population. There are
several advantages in studying this galaxy population in the
infrared: the NIR bands sample galaxy rest-frame visible spectral
energy distributions (SEDs) for galaxies at $z\sim2$  and thus trace
their stellar mass better than rest 
UV observations can \citep{cowie1994}, and  the mid-IR (MIR)/far-IR
(FIR) bands permit measurement of star formation even in very dusty
galaxies.

Existing deep NIR surveys have identified massive, passive galaxies
already in place at $z\sim2$, implying that such galaxies formed at even
higher redshifts \citep{franx2003, glazebrook2004, mccarthy2004,
  labbe2005, daddi2005}.    Based on study of
local massive galaxies, the current theoretical view is that
massive galaxies are formed through major mergers \citep{cole2000,
  naab2003, kormendy2009}. A major merger also triggers the
intensive star formation phase known as  Ultra-Luminous InfraRed
Galaxies (ULIRGs) and feeds gas to central massive black holes
to create QSOs \citep{sanders1988, veilleux2009,
  hou2011}). ULIRGs at $z\sim2$ have been detected by a variety of
methods and are known by many names including SubMillimeter Galaxies
(SMGs), MIPS 24~\micron\  
selected ULIRGs, and Dusty Obscured Galaxies (DOGs)
\citep{chapman2003, houck2005, yan2007, huang2009,
  desai2009}. However, the existence of so many massive galaxies
with stellar masses $M_*>10^{11}$~\Msun\ at high redshifts challenges
the merger 
scenario for the formation of massive galaxies.
Current numerical simulations \citep{narayanan2009}
have failed to produce as many major mergers as required to
explain the observed number of ULIRGs at $z\sim2$. \citet{davé2010}
proposed an alternative formation scenario for SMGs: a massive gas-rich
galaxy could have star formation rate (SFR) as high as
180--500~\Msun~yr$^{-1}$ without any merging process, implying 
that ULIRGs at $z\sim2$ with $\LIR$ just above $10^{12}$~\Lsun\
may have a different formation mechanism than more luminous
ULIRGs. A complete census of 
ULIRGs over the full luminosity range is needed to solve the puzzle
of massive galaxy formation at $z\sim2$. 

Morphological studies can elucidate the ULIRG formation process by showing the
presence or absence of merger signatures.  Recent morphological studies
\citep{dasyra2008, melbourne2008, melbourne2009, 
  bussmann2009, bussmann2011, zamojski2011, kartaltepe2012} using
high-angular-resolution NIR images have shown that $z\ga2$
ULIRGs exhibit a  wide range of forms from unresolved to complex
structures and sometimes but not always multiple components.
The diversity of morphologies indicates that ULIRGs may occur in different
interaction stages of major mergers, in minor mergers, or via secular
evolution not involving mergers at all. 

Most ULIRGs are so optically faint that they were recognized as a class
only when infrared satellite surveys became available.
The optical faintness is because the UV-optical emission 
is absorbed by  dust and re-emitted in the far-infrared \citep{sanders1996}. 
Therefore infrared  surveys are needed to give unbiased samples of
ULIRGs.  Redshift surveys of SMGs
\citep{chapman2003}  have revealed a much larger  ULIRG
population at $1.7<z<2.8$ than at $z\approx0$.  SMGs, 
however, represent a selection method  known to be biased toward low dust
temperature 
systems. In contrast, \citet{magdis2010} showed that selecting a sample based on
MIPS 24~$\mu$m sources with IRAC 3.6--8~\micron\ colors indicating
that the 1.6~\micron\ stellar 
``bump'' is near  4.5~\micron\ \citep{huang2009} produced a $z\sim2$
ULIRG sample having a wide range of dust temperatures.

This paper defines a lower luminosity sample of $z\approx2$ ULIRGS
and examines their properties, including using {\it HST}/WFC3 NIR 
images to examine morphologies.  The study is enabled by
the data release of the EGS region of the Cosmic Assembly Near-IR
Deep Extragalactic Legacy Survey \citep[CANDELS:][]{grogin2011,
koekemoer2011}. The CANDELS survey covered $\sim$210~arcmin$^2$ with
a total exposure time in the EGS of 90 HST orbits.  The survey reaches
$5\sigma$ point-source depth of $H_{\rm AB}\sim 26.5$ at
resolution (measured from PSF FWHM) of 0\farcs12--0\farcs18.  The NIR
data are complemented by $I$ (F814W) data having resolution
0\farcs08--0\farcs09 from {\it HST}/ACS\null. (Pixel scales are
0\farcs03/pixel in ACS and 0\farcs06/pixel in WFC3.)

Infrared spectroscopic ULIRG surveys are particularly valuable.
In addition to giving redshifts, MIR spectra of ULIRGs divide the
objects into two types.  Objects with strong power-law continua are
powered mainly by active galactic nuclei (AGNs)
\citep{Roche1984,houck2005, sajina2007, weedman2006, dasyra2009}, while those
with strong polycyclic aromatic hydrocarbon (PAH) emission are
powered by intensive star formation \citep{Roche1985,weedman2006,
  farrah2008, dasyra2009, desai2009, huang2009, fadda2010, fiolet2010}.  At
$z\sim2$, the MIPS 24~\mum\ band probes rest-frame $\lambda\sim
8~\mum$ where there is a strong PAH emission feature. This means
samples selected at 24~\micron\ will be especially effective in
finding star-forming ULIRGs.

Most work on $z\sim2$ ULIRGs has concentrated on objects with
$L_{\rm IR} > 10^{12.5}$ \citep{houck2005, weedman2006, yan2007,
  farrah2008, desai2009, huang2009, fiolet2010}. 
This paper is instead a
study of a lower-luminosity sample of 14 ULIRGs with $10^{12.0} < L_{\rm
  IR} < 10^{12.6}$. Section 2 of the paper describes the sample
selection.  The IRS spectroscopic results are presented in Section 3,
and Section 4  analyzes stellar populations,  SFRs,
stellar masses ($M_*$), total infrared luminosities
(\LIR), morphologies, and AGN fraction.  Finally, a brief summary is
presented in Section 5. All magnitudes and colors are in the AB
system, and notation such as ``[3.6]'' means
the AB magnitude at wavelength 3.6~\micron. The paper
uses cosmological parameters $h\equiv H_{\rm 0}$[km s$^{-1}$
  Mpc$^{-1}$]$/100=0.71$, $\Omega_\Lambda =0.73$, $\Omega_{\rm M} =
0.27$.

\section{Sample Selection}\label{sec:sam}

For the present study, IRS targets were selected from the
24~\micron\ sources \citep{papovich2004} in the Extended Groth Strip
(EGS) region, where the survey was more than 80\% complete for $\FMI >
0.11$~mJy.  \citet{huang2009} presented IRS spectroscopy for a 
24~\micron\ sample with $\FMI >0.6$~mJy and obtained a 
narrow redshift distribution at $z\sim1.9$. For this paper, we
selected a fainter sample of 14 objects with $0.2<\FMI<0.6$~mJy. Most
objects in our sample have $0.2<\FMI<0.5$~mJy with only two having
$0.5<\FMI<0.6$~mJy.

\begin{figure}
\centering
\includegraphics[angle=0,width=\columnwidth]{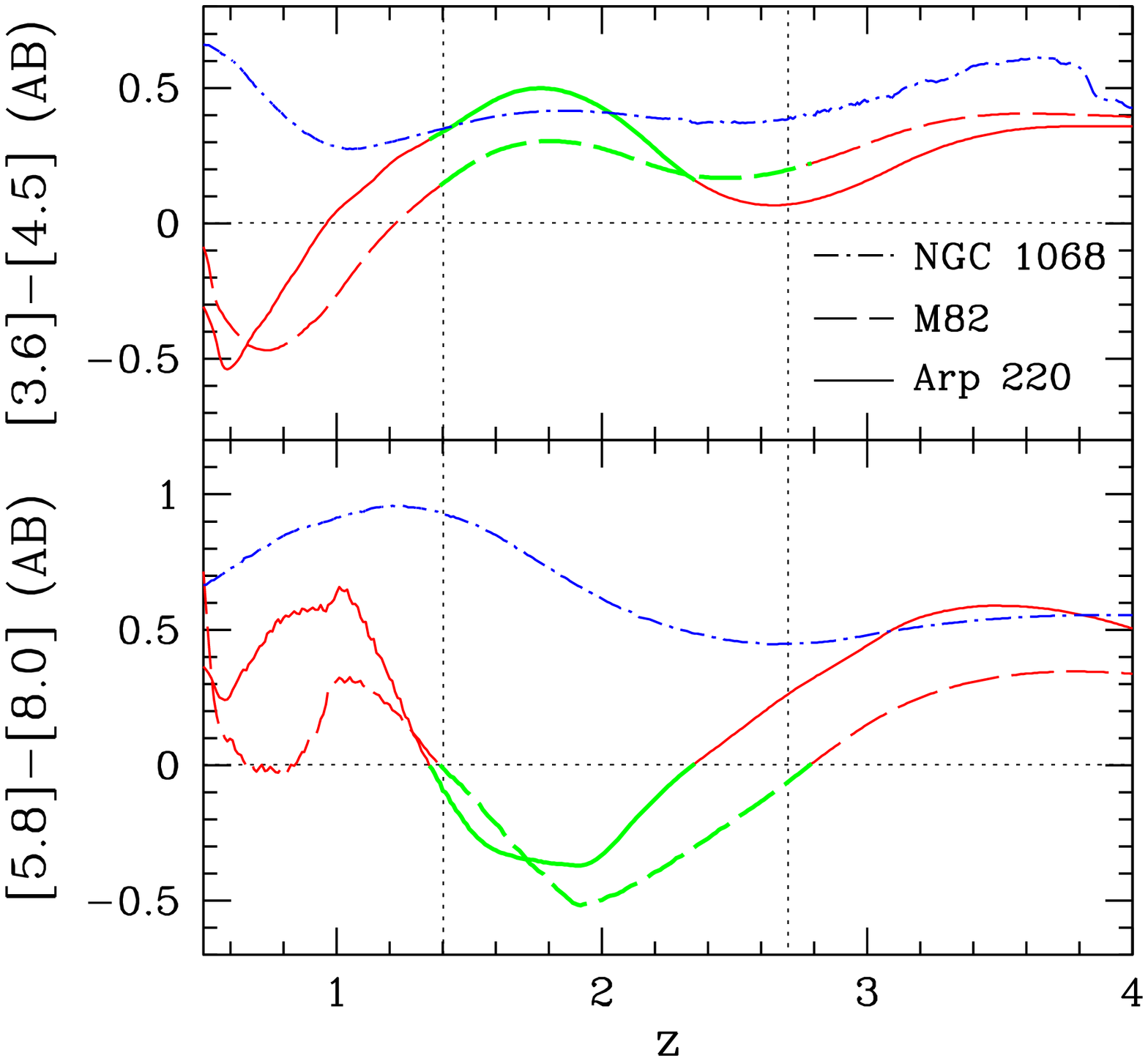}
\caption{ IRAC selection criteria for $z\sim2$ ULIRGs. Lines show
  expected IRAC colors for three local template sources as a function
  of redshift. M82 (dashed lines) is a starburst, Arp~220 (solid
  lines) is a ULIRG, and NGC~1068 (dot-dashed lines) is an AGN\null.
  Lines are plotted in green where each template would have been
  selected according to the color criteria of inequality~1, which are
  shown as horizontal dotted lines. A galaxy with the NGC~1068 SED
  would not be selected at any redshift. Vertical dotted lines
  correspond to redshifts 1.4 and 2.7 respectively and show the
  approximate range of color selection.
\label{fig:model}}
\end{figure}

In the redshift range of $1.4<z<2.7$,
the four IRAC bands (3.6--8.0~\micron) probe the rest-frame NIR bands
where nearly all galaxy stellar population SEDs 
have similar shapes.  In particular, the rest 1.6~\mum\ stellar emission peak is
nearly always present independent of redshift or metallicity.  As described by \citet{huang2009}, at $z<1.4$, the 
IRAC 3.6 and 4.5~\mum\ bands sample the Rayleigh-Jeans tail of stellar emission
resulting in $[3.6]-[4.5]<0$, while at $z>1.4$, the 1.6~\mum\ bump moves 
beyond the IRAC 3.6~\mum\ band, and  $[3.6]-[4.5]>0$. These 
properties make color selection with IRAC 
comparable to BzK \citep{daddi2004} selection in completeness and
contamination \citep{sorba2010}, and 
\citet{huang2004} and \citet{papovich2008} have used IRAC color
to select galaxies at $z>1.4$. A limitation of the technique is that
very dusty galaxies  can have 
$[3.6]-[4.5]>0$ even at $z<1.4$. We therefore propose to use $[5.8]-[8.0]<0$ to exclude dusty
galaxies and also AGNs with power-law-like SEDs. This cut also excludes
galaxies at $z\ga2.7$ because the 1.6~\micron\ bump begins to move past the
5.8~\micron\ band.  Therefore the adopted color
criteria for our faint 24~\micron\ sample for IRS spectroscopy are:
\begin{equation}
[3.6]-[4.5]>0~~~\&~~~[5.8]-[8.0]<0.
\end{equation}
These color criteria select galaxies in the redshift range of $1.4
\la z \la 2.7$ as illustrated in
Figure~\ref{fig:model}. 
Throughout this redshift range,  the 7.7~\mum\ PAH emission feature is within the wavelength
coverage of the IRS, enabling redshift measurements. 
\citet{huang2009} used a similar selection
\begin{equation}
0.05< [3.6]-[4.5]< 0.4~~~\&~~~-0.7< [3.6]-[8.0]<0.5 ,
\end{equation}
but using a fourth IRAC wavelength (5.8~\micron) as in criteria (1) does
a better job of rejecting galaxies with near-power-law SEDs because
all power laws lie outside the criteria (1) selection while some
(those with ($-0.6\la\alpha\la -0.2$ for $F_\nu\propto\nu^\alpha$)
lie inside criteria (2).

\begin{figure*}
\centering
\includegraphics[angle=0,width=0.9\textwidth]{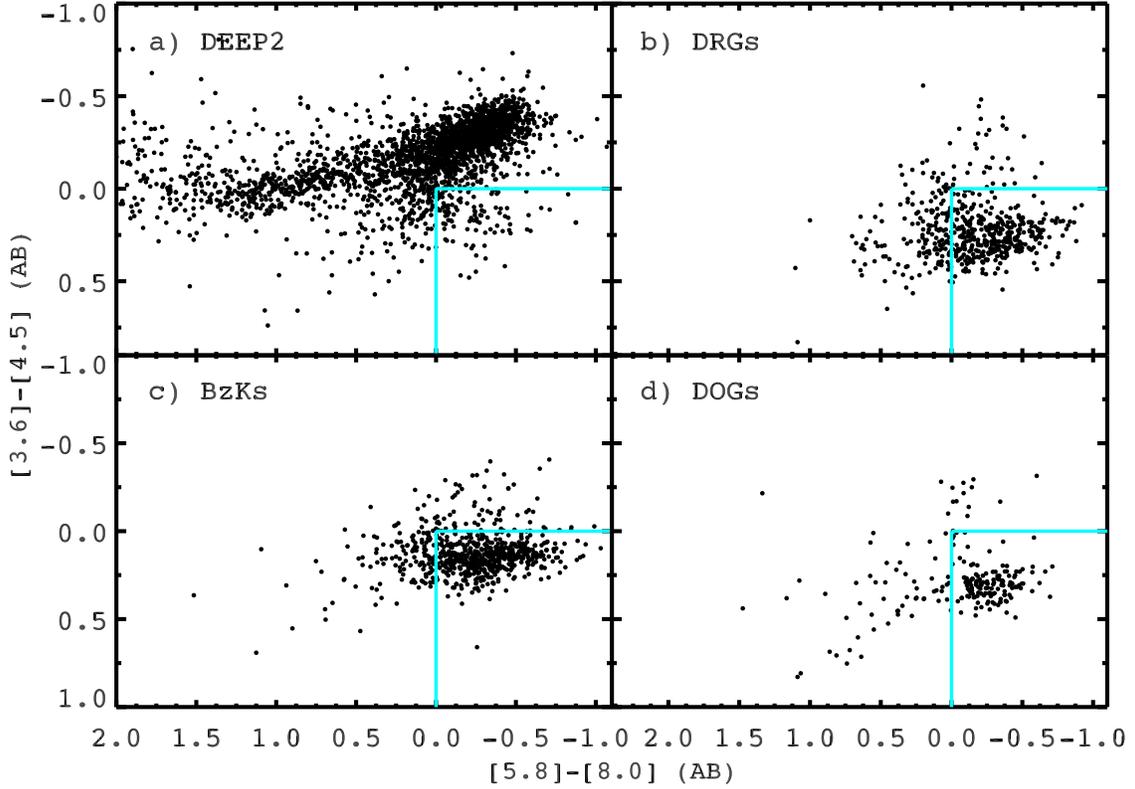}
\caption{IRAC color-color diagrams for observed galaxy samples. 
a) AEGIS spectroscopic redshift $z<1.4$ sample (with redshift quality
$\rm Q\ge3$. $\rm Q=3$: Secure redshift, $\rm Q=4$:
Very secure redshift);
b) Distant Red Galaxies \citep[DRGs;][]{franx2003} in AEGIS; 
c) BzK galaxies \citep{daddi2004} in AEGIS; d) DOGs \citep{dey2008} in AEGIS.
Cyan lines show the  IRAC color criteria (1).  
\label{fig:color}}
\end{figure*}

\begin{figure*}
\centering
\includegraphics[angle=0,width=0.9\textwidth]{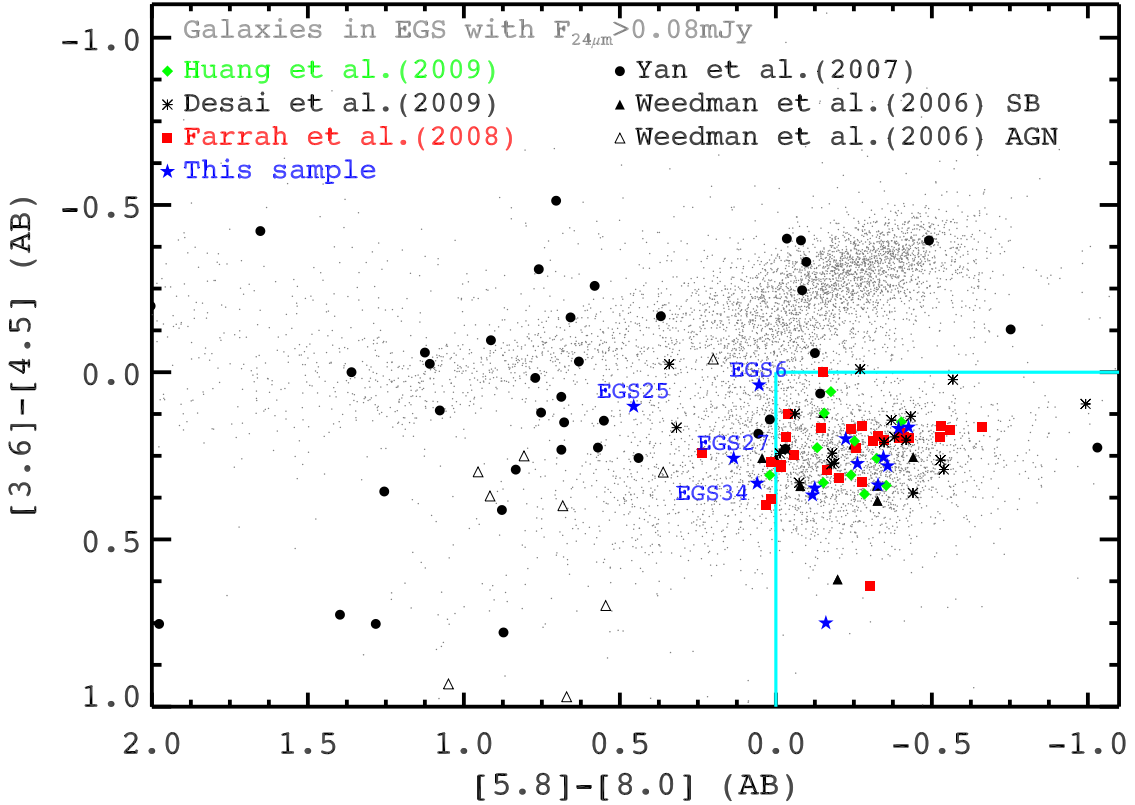}
\caption{ IRAC color-color diagram for EGS galaxies with
  $\FMI>0.08$~mJy.  Small dots show all such galaxies; blue stars
  show galaxies in the current IRS spectroscopic sample, which also
  requires $\FMI>0.2$~mJy.  The cyan line shows the IRAC color
  criteria (1). Labels indicate  EGS6, which slightly misses criteria
  (1) because of the initial
  the photometry error, and  three X-ray objects that
  are AGN candidates. Objects from other IRS spectroscopic samples (Table~\ref{tab2}) at $z\sim2$
are plotted for comparison, $90\%$ of all sources in these SB-dominated 
samples reside in our IRAC color region \citep{weedman2006, farrah2008, desai2009, huang2009}.
\citet{yan2007} used extreme optical-to-24 \mum\ color to select 
dusty sources. Sources in this sample have much redder  
$[5.8]-[8.0]$ IRAC colors than the majority of 24 \mum\ sources 
and are mostly AGNs as shown by their strong
power-law continua, but weak or absence of PAH emission features.
\label{fig:criteria}}
\end{figure*}

Figure~\ref{fig:color} compares criteria (1) with
other selection methods, which all effectively select
galaxies at $z\sim2$.  However, only the IRAC selection uses the
rest-frame NIR bands, making the sample selection nearly unaffected by dust
reddening.  Moreover, NIR emission is closely tied to stellar mass
\citep{Bell2001}, 
and the resulting sample is therefore roughly equivalent to a
stellar-mass-selected sample.  It is thus ideal for studying
luminous, massive galaxies \citep{huang2004,
  conselice2007}.
As expected, 24~\micron\  sources show a range of IRAC colors, but
Figure~\ref{fig:criteria} shows that one  dense
concentration is  in the region cornered by 
criteria (1).  The present sample consists of  11 objects with
$0.2<\FMI<0.6$~mJy and satisfying criteria 
(1)\footnote{One object, EGS6, was initially selected with criteria (1) and 
   observed with IRS\null. After the IRS observation, the EGS
  IRAC images and catalog were updated with new IRAC imaging
  from the \s\ GO program 49888 (PI:
  Nandra). The updated color for EGS6 became
  $[5.8]-[8.0]=0.05$, slightly too red to qualify per criteria (1),
  but we include this source  nevertheless.} and 3 X-ray sources 
(EGS25/EGS27/EGS34) with colors satisfying the criteria in
\citet{huang2009} for comparison.  The
positions and flux densities for objects in the sample are listed in
Table~\ref{tab1}.

There are many IRS surveys for IR luminous sources at $z\sim2$ using
samples with differing criteria \citep{weedman2006, yan2007,
  farrah2008, desai2009, huang2009}.  When color selection is used,
there are two general
categories of  selection criteria: DOGs selected by very red
visible to infrared colors 
and galaxies selected in the infrared with SEDs peaking
at $\ga$4.5~\micron.  Table~\ref{tab2} summarizes selection criteria
for the relevant IRS surveys, and Figure~\ref{fig:criteria} shows how
the various samples compare with criteria (1).  \citet{houck2005}
used a selection of the first type: MIPS 24~\micron-luminous sources
very faint at visible wavelengths. Objects of this type show strong
MIR continua with $[5.8]-[8.0]$ colors much redder than the majority
of 24~\mum\ sources, strong silicate absorption, and weak or absent
PAH emission features.  Similar types of objects observed by other
groups \citep{weedman2006, yan2007} yield similar results.  The
strong power law continua and weak or absent PAH emission features
show that these sources are AGNs. However, fainter sources (median
$\FMI=0.18$~mJy) selected
by the DOG criterion \citep{pope2008} are predominantly star
forming.  Selection via IRAC colors
\citep{weedman2006, farrah2008, desai2009, huang2009} finds
objects whose MIR spectra show strong PAH emission features,
indicating that star formation powers their FIR emission. When
all existing spectra are used to categorize sources, 90\% of SB-dominated
objects meet criteria (1), though this result is biased by the
initial sample selections.

\section{IRS Observations and Data Reduction}\label{sec:irs}

IRS observations of our sample were made as part of the GTO program for the
\s/IRAC instrument team (program ID:\ 30327). Objects were observed
with the IRS Long-wavelength Low-resolution first order (LL1)
mode with wavelength coverage 20--38~\mum\ and slit width 10\farcs7.
For galaxies at 
$z\sim2$, major spectral features including the PAH emission
features at 7.7, 8.6, and 11.3~\mum\ and silicate absorption from
8 to 13~\mum\ (peaking at 9.7~\mum) fall in the  observable wavelength
range.  IRS
observation of each object consisted of 6 exposures with ramp
duration 120~s. Mapping mode
\citep{teplitz2007} was used, offsetting the  pointing by 24\arcsec\ along
the IRS slit between exposures.  This  mode not
only gives more uniform spectra for the targets but also better rejects
cosmic rays and bad pixels.  All spectra were  processed initially with
the \s\ Science Center pipeline version 13.0. Extraction of source
spectra was done with both the {\tt SMART} analysis package
\citep{higdon2004} and customized software \citep{huang2009} to
produce calibrated spectra.

\begin{figure*}
\centering
\includegraphics[angle=0,width=0.9\textwidth]{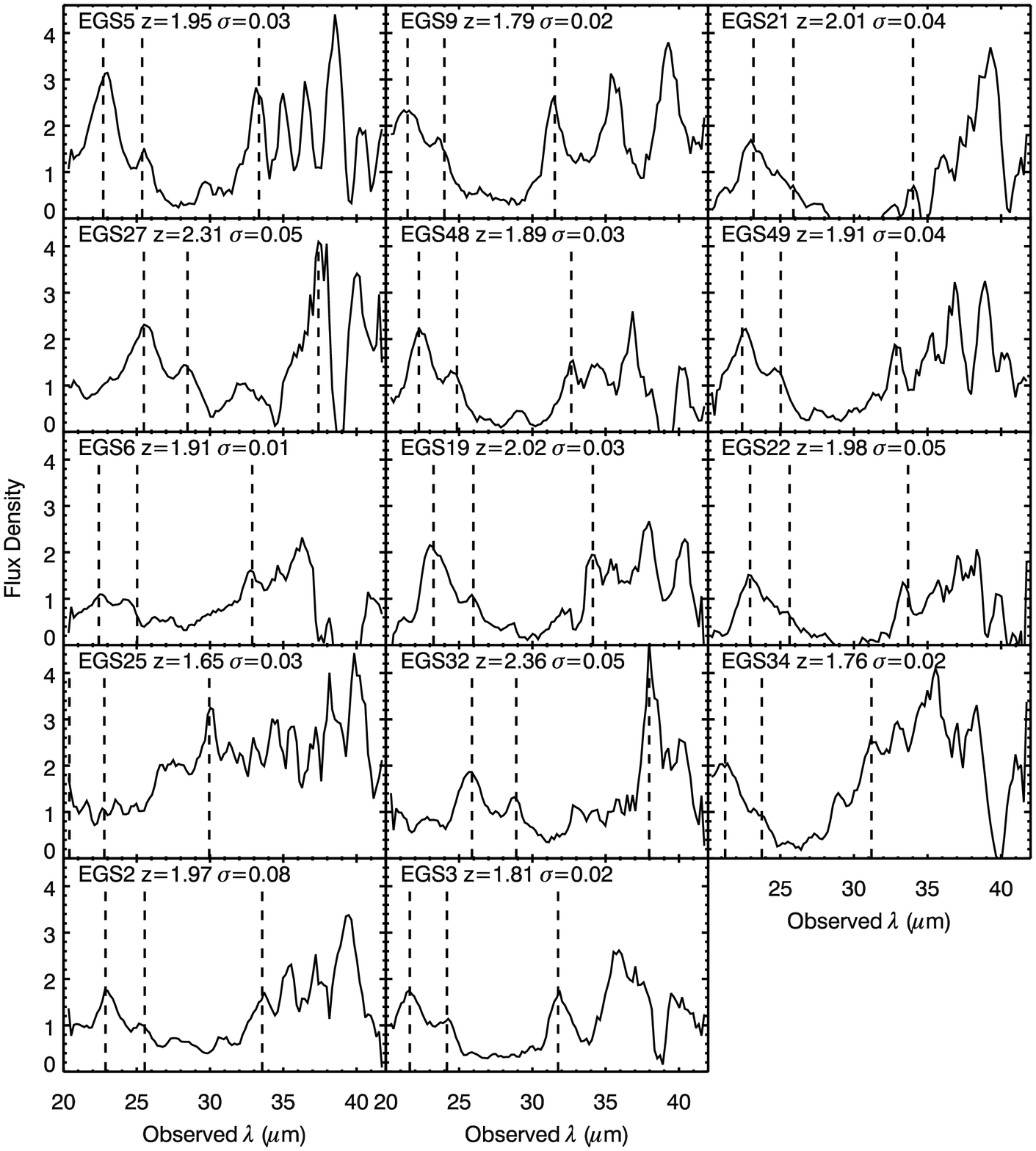}
\caption{Mid-IR spectra for the 14 sources observed with
  IRS\null.  The vertical scales are linear but arbitrary and different for each 
  panel. The spectra were smoothed by a four-pixel boxcar
  in order to enhance the broad features such as PAH emission and
  silicate absorption.  Dashed lines indicate the central wavelengths of the
  PAH emission features at rest-frame 7.7, 8.6, and 11.3~\mum\ from
  left to right. The source nicknames, redshifts, and
  redshift uncertainties derived from the template fit are shown in each panel.
  \label{fig:irs}}
\end{figure*}

\begin{figure}
\centering
\includegraphics[angle=0,width=0.6\textwidth]{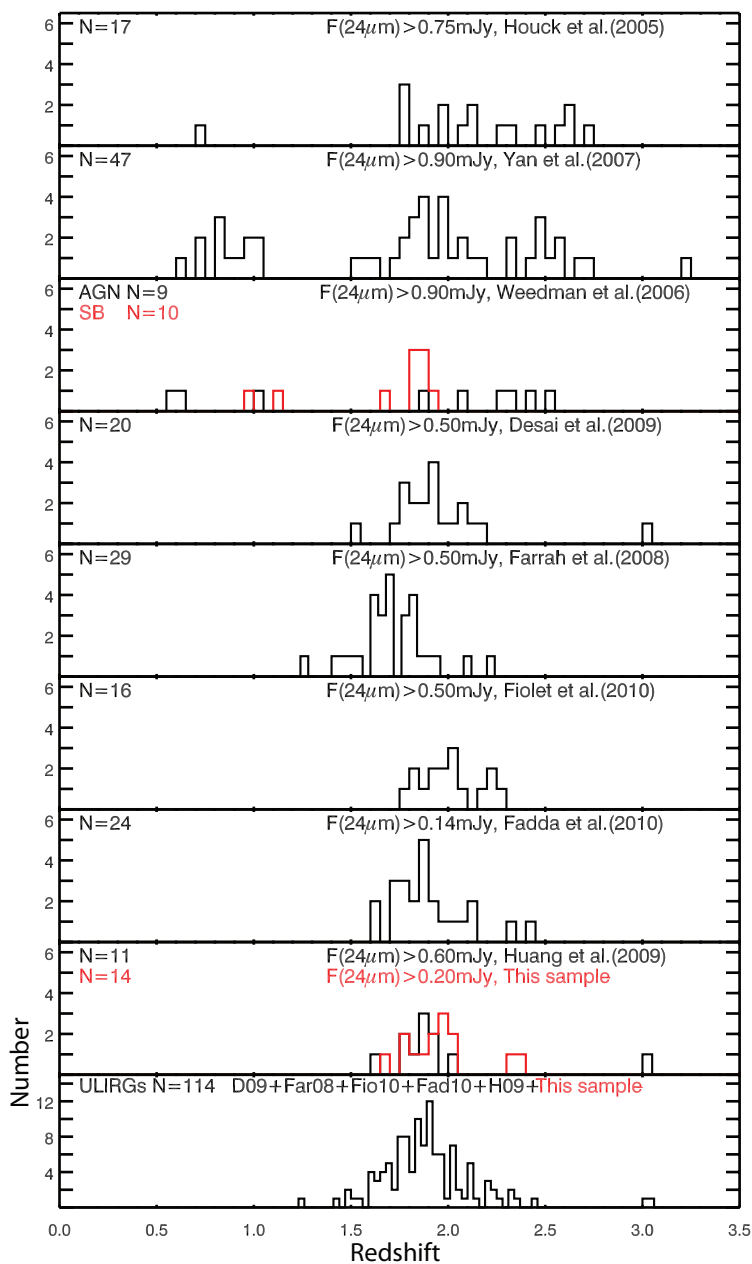}
\caption{ Redshift distributions for the IRS spectroscopic samples in
  Table~\ref{tab2}. The number  of objects in each sample and
  their source are indicated in each panel.
  AGNs and starbursts from \citet{weedman2006} are shown separately.  
 The bottom panel shows the redshift distribution for
  all 114 SB-dominated sources in the combined samples 
(D09+Far08+Fio10+Fad10+H09+ours, see Table~\ref{tab2} for
detail).
\label{fig:specz}}
\end{figure}

Figure~\ref{fig:irs} shows the IRS spectra of the 14 sources in our
sample. All objects in the sample show PAH emission features at 7.7,
8.6, 11.3~\mum\ in their spectra, and some have silicate absorption
at 9.7 \mum. We measured redshifts by
cross-correlating the observed spectra with
two local templates, M82 and Arp~220. The two templates
yield very nearly the same redshifts with a typical difference of $\Delta
z=0.02$. The M82 template fits all spectra better. Figure~\ref{fig:specz} shows
the redshift distribution of our sample compared with some works from the 
literature. Our sample all lies within $1.6<z<2.4$ (Table~\ref{tab3}), 
demonstrating the efficiency of our selection criteria. The three
X-ray sources have spectra generally  
similar to the rest of the sample.  EGS25 and EGS27 have weaker silicate
absorption and higher continuum than most sources,  consistent with their
red $[5.8]-[8.0]$ colors in
Figure~\ref{fig:criteria}. EGS34, however, has strong silicate
absorption and weak PAH emission.

Figure~\ref{fig:specz} shows that the redshift distribution of our
sample is very similar to that of the SB-dominated ULIRGs selected
with higher limiting flux densities \citep{farrah2008, desai2009,
  huang2009, fadda2010, fiolet2010}. The narrow redshift distribution
for the ULIRGs is due to the selection of strong 7.7~\mum\ PAH
emission by the MIPS 24~\mum\ band at $z\sim1.9$.  The mean redshift
of 114 ULIRGs in the various surveys is $\langle z\rangle =1.89$ with
a dispersion $\sigma=0.25$. Our sample has $\langle z\rangle = 1.95$
and $\sigma= 0.19$. On the other hand, luminous
24~\mum\ sources with power-law SEDs have a much wider redshift range
extending from $z\sim0.5$ to $z\sim3$ \citep{houck2005, weedman2006,
  yan2007}.

\section{Multi-Wavelength Studies of ULIRGs at z$\sim$2}\label{sec:resu}

\begin{figure*}
\centering
\includegraphics[angle=0,width=0.9\textwidth]{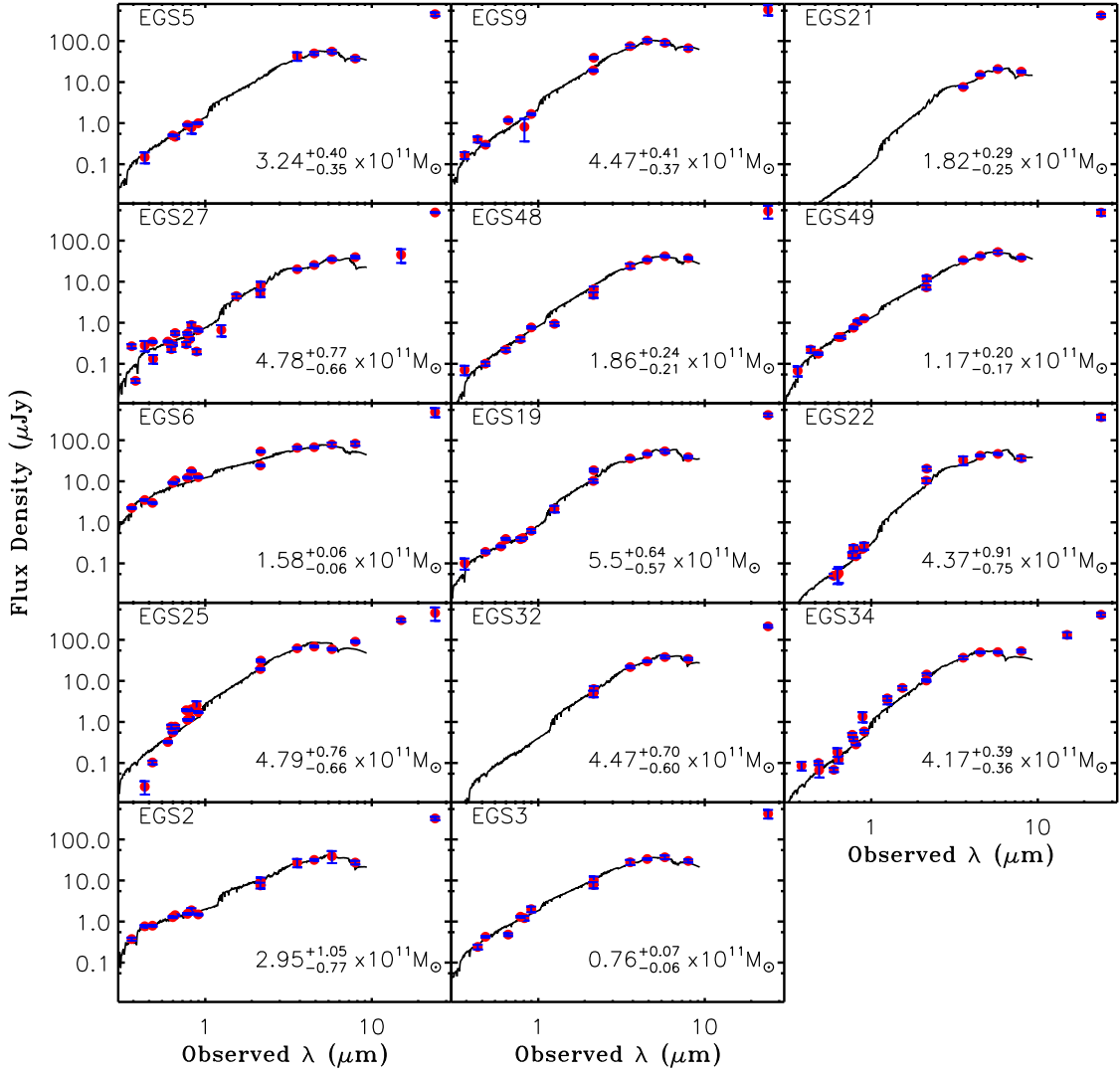}
\caption{Spectral energy distributions for the 14 sample
  galaxies.  Points show the observations, and  black lines show the
  best-fit  stellar population model (CB07). The source nickname and
  inferred stellar mass ($M_*$) are given in each panel.
\label{fig:fit}}
\end{figure*}

AEGIS (All-wavelength Extended Groth Strip International Survey) is a
multi-wavelength survey covering  X-ray to FIR bands in the
Extended Groth Strip area \citep{davis2007}.\footnote{AEGIS data
  products are described at http://aegis.ucolick.org/astronomers.html
  , and data are included in the ``Rainbow'' data compilation at
  https://rainbowx.fis.ucm.es/Rainbow\_Database/Home.html .} The rich
multi-wavelength data permit the study of SEDs and physical
properties for objects in our sample. MIR and FIR
photometry for this sample is particularly important in determining
their properties. All but the three X-ray sources were detected by \AKARI\ at
15~\mum.  One object (X-ray source EGS34) was detected at 70~\mum\ in
the FIDEL survey.  The three X-ray-selected objects and EGS22
were detected in the {\it Chandra} 800~ks X-ray imaging
\citep{laird2009}.  Four objects in our sample, EGS6/EGS9/EGS25/EGS34, are in
the VLA 1.4~GHz radio catalog  \citep{ivison2007,willner2012}.
Figure~\ref{fig:fit} shows SEDs of the sample galaxies.

\subsection{Total Infrared Luminosity and Star Formation 
Rate}\label{sec:sfr}

FIR luminosity is an important measurement in characterizing
ULIRGs at $z\sim2$. ULIRGs with different $\LIR$\footnote{We adopt
  the Sanders \& Mirabel (1996) definition of $\LIR \equiv L{(8\,\micron
    -1{\rm\, mm}}$).} and thus different SFRs may have undergone different formation
processes. ULIRGs may be the dominant contribution to star
formation density at $z\sim2$ \citep{lagache2004, caputi2007}, making
them especially important to characterize. Direct measurement of
$\LIR$ requires FIR (${\sim}100$~\micron) photometry, which can sample the 
peak of the dust emission SED\null, but FIR photometry is not yet available for most of the
sources.  Many groups have
made substantial efforts to convert MIR luminosities into $\LIR$
\citep{chary2001, reddy2006, caputi2007, bavouzet2008,
  rodighiero2010, wu2010}, but the results remain uncertain.

In order to estimate SFR,
we derived $L(8~\micron)$ for our sample from the observed 
24~\mum\ flux densities.  At $z\sim2$,  24~\mum\ corresponds to
rest-frame 8~\mum\ with its strong PAH emission feature. 
Because of the strong emission feature and its difference from object
to object,
the K-correction may have large
scatter.  Objects in our sample are in a rather
narrow redshift range, and the  K-correction needed to covert
24~\mum\ flux to the $L(8~\mum)$ is close to zero. We stacked
all the
spectra together to generate a mean rest-frame spectrum for the sample and used
it to calculate the K-correction as a function of redshift. The
derived 8~\mum\ luminosities are given in Table~\ref{tab3} and range
from $10^{11.41}$ to $10^{11.79}$~\Lsun. 

\begin{figure}
\centering
\includegraphics[angle=0,width=\columnwidth]{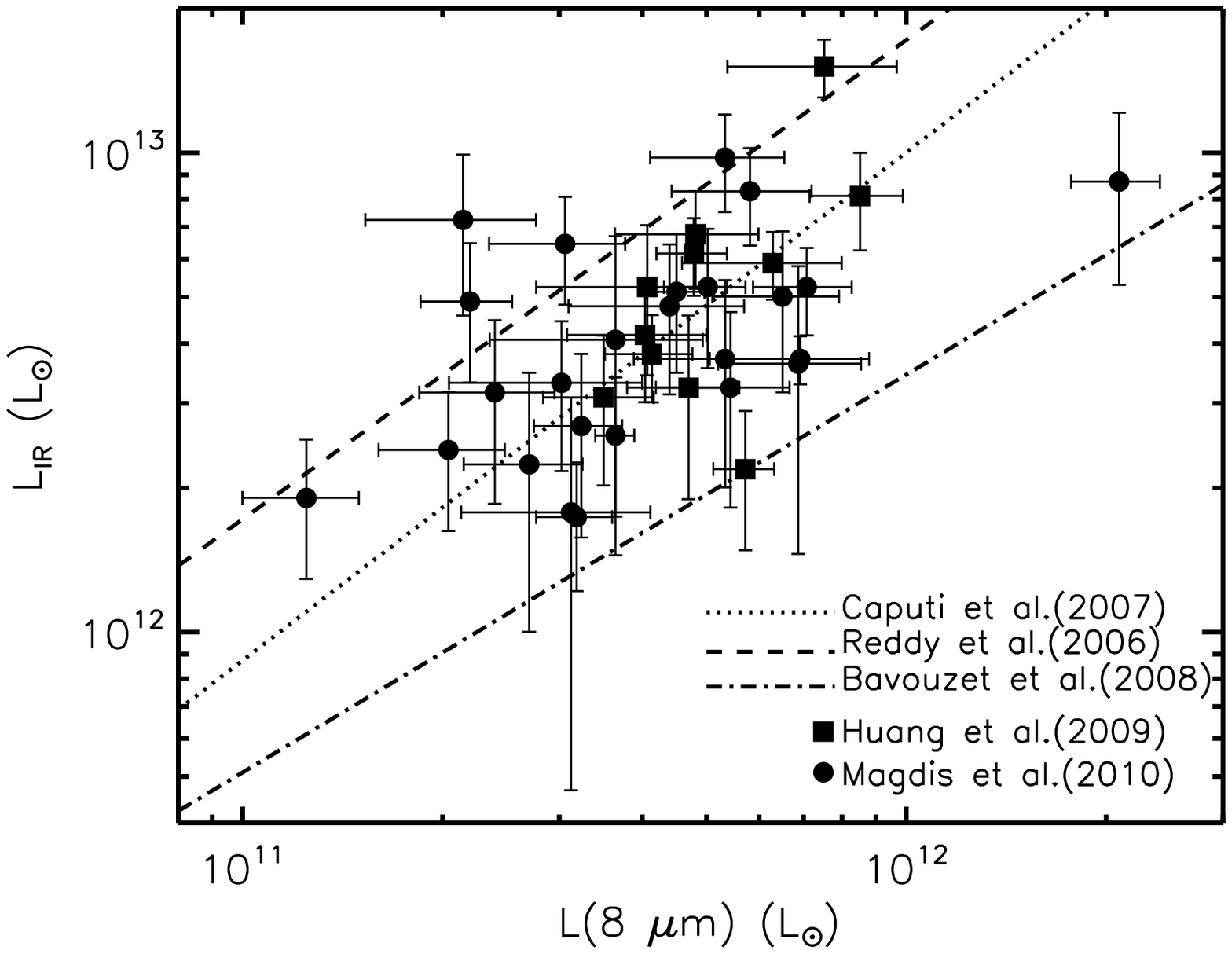}
\caption{Observed IR luminosity  versus rest-frame 8~\micron\
  luminosity ($\nu L_\nu$) for 36 individual ULIRGs at
  $z\sim2$. Filled squares denote data from \citet{huang2009} and
  filled circles from   \citet{magdis2010}. Lines show the local
  relations found by
  \citet{caputi2007} (dotted) for 24~\micron-selected galaxies and by
  \citet{bavouzet2008} (dot-dashed) for FIR-selected galaxies.  The
  dashed line shows the   relation found by 
  \citet{reddy2006} for $z\sim2$  galaxies
  selected by observed visible--UV colors.
\label{fig:lir}}
\end{figure}

There have been several studies of the $L(8~\mum)$--$\LIR$ relation.
\citet{bavouzet2008} used an IR-selected galaxy
sample at $z<0.6$ to derive an empirical relation
$\LIR=1.3\times10^{12}(L(8~\mum)/10^{11.5})^{0.83}$. \citet{caputi2007}
used the same 
sample to find 
$\LIR=3.0\times10^{12}(L(8~\mum)/10^{11.5})^{1.06}$, i.e., slightly
higher $\LIR$ for a given $L(8~\mum)$. The
inconsistency may come from 
objects at $0.3<z<0.6$ in the sample. \citet{huang2007} showed that
the 7.7~\mum\ PAH feature begins to shift out of the IRAC 8 \mum\ band for
$z>0.3$. The K-correction to calculate $L(8~\mum)$ for
galaxies in $0.3<z<0.6$ from their 8~\mum\ flux densities is strongly
model dependent and  may introduce a large uncertainty in the
resulting $L(8~\mum)$. On the other hand, the $L(8~\mum)$--$\LIR$
relation at different redshifts may  differ because IR
samples select different populations at different
redshifts. \citet{sajina2008} performed MIPS 70 and 160~\mum\ and
ground-based millimeter imaging of   $z\sim2$ ULIRGs and
measured $\LIR$ directly from the FIR and millimeter
photometry. Their $L(8~\mum)$--$\LIR$ relation
is consistent with that of \citet{bavouzet2008}.
\citet{huang2009} performed FIR and millimeter photometry for their
ULIRG sample and obtained a different $L(8~\mum)$--$\LIR$ relation
as shown in Figure~\ref{fig:lir}. This relation was recently confirmed by
\citet{magdis2010} with \h/PACS and SPIRE imaging  at 100,
160, 250, 350, and 500~\mum. Figure~\ref{fig:lir} shows that, for a
given $L(8~\mum)$, FIR-selected galaxies
\citep{caputi2007,bavouzet2008} appear to have higher $\LIR$ than  galaxies
selected by UV--visible color \citep{reddy2006}, though this may be
confounded by the differing sample redshifts.
Figure~\ref{fig:lir} also shows that  the samples of both
\citet{huang2009} and \citet{magdis2010} have an
$L(8~\mum)$--$\LIR$ relation consistent with that of
\citet{caputi2007}. The galaxies in our sample have lower $L(8~\mum)$
than those of \citet{huang2009} and \citet{magdis2010}, and their
$L(8~\mum)$--$\LIR$ relation is unknown.
We have used the \citet{caputi2007} relation to calculate $\LIR$, but
the results  are yet to be confirmed with \h/SPIRE
photometry. Table~\ref{tab3} includes $\LIR$ 
calculated with both the  \citeauthor{caputi2007} and the
\citet{bavouzet2008} relations for comparison. All sources in the sample have
$L_{\rm IR} > 10^{12}$~\Lsun\ no matter which relation is used. 
Table~\ref{tab3} also shows the derived SFR using the Kennicutt
(1998) relation\footnote{ ${\rm 
  SFR}\,(M_\odot\ {\rm yr}^{-1})=4.5\times10^{-44} \LIR~({\rm erg}\,
{\rm s}^{-1})$} applied to the \citeauthor{caputi2007} $\LIR$.
The  median SFR for ULIRGs in our sample is 570~\Msun~yr$^{-1}$
(250~\Msun~yr$^{-1}$ with the \citeauthor{bavouzet2008}  relation).

\subsection{Stellar Population and Mass in ULIRGs}\label{sec:mass}

Stellar population modeling  \citep[\eg,][]{bruzual2003} provides a
way to determine stellar
parameters from  observed photometry.  
ULIRGs have a bursty star formation history, very young stellar
populations, and non-uniform dust distribution, all of which 
introduce large uncertainties in modeling their stellar
populations. Despite those concerns, stellar masses deduced from rest NIR data
are the most 
robust property against variations in star formation history,
metallicities, and the extinction law \citep{forster2004}. In deriving stellar masses, we
assumed constant SFR, which should be a good approximation given that
the ULIRGs are observed to be undergoing intensive star 
formation. Several groups have demonstrated that a constant SFR
provides a reasonable description of stellar population
evolution for galaxies with ongoing star formation at high redshifts,
such as LBGs, Lyman-alpha emitters (LAEs), star-forming BzKs, and
DRGs \citep{shapley2001, van2004, rigopoulou2006, kong2006,
  lai2007}. With very young stellar populations,  stars on the
asymptotic giant branch (AGB) make
a significant contribution to the galaxy NIR emission
\citep{maraston2005}. The initial mass function (IMF) is also in
question; \citet{daddi2007} argued that the Kroupa IMF 
fits  ULIRGs beter than other choices. For the present work, we fit
the observed SED of each source using updated \citet{bruzual2003}
models  (S.\ Charlot 2006 private communication, but widely known as
CB07) with a Kroupa IMF and a constant star formation rate.
~Figure~\ref{fig:fit} shows the
observed photometry,  best-fit models, and inferred stellar masses
for the 14 ULIRGs in our sample. The stellar masses  are in the range
$10.9< \log(M_*/M_{\odot})<11.7$.

\begin{figure}
\centering
\includegraphics[angle=0,width=\columnwidth]{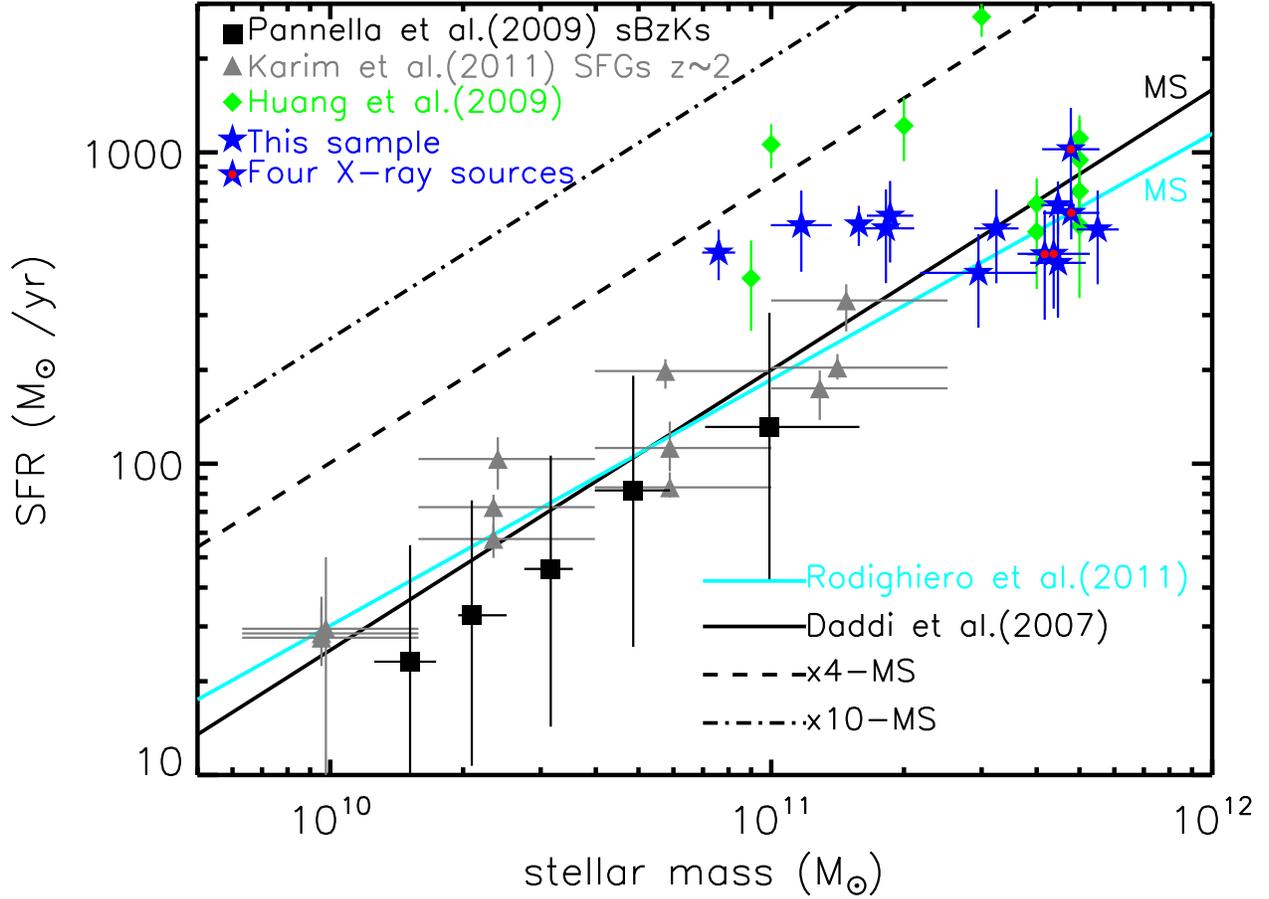}
\caption{ Stellar mass--SFR relation at $z\sim2$. Blue stars denote
  objects in the present sample with X-ray sources indicated by
  superposed red dots.  Filled diamonds denote individual ULIRGs
  from \citet{huang2009}.  Filled triangles denote average SFRs for
  star-forming  galaxies in different mass bins
  \citep{karim2011}; for each mass bin, three different redshift bins
  from $z=1.6$ to 3.0 are plotted.  Filled squares denote average
  SFRs for BzK galaxies with AGNs excluded
 \citep{pannella2009}.
  Solid black and cyan lines indicate the main sequence (MS) for
  star-forming galaxies at $1.5<z<2.5$ as defined by
  \citet{daddi2007} and \citet{rodighiero2011}, respectively. 
  Dot-dashed and dashed lines mark the loci 10 and 4 times above the
  \citeauthor{daddi2007} MS.
\label{fig:ms}}
\end{figure}

Star-forming galaxies in the local universe follow a tight
correlation between 
stellar mass and SFR, defining a main sequence (MS)
\citep{brinchmann2004, peng2010}.  The MS is also seen
at  $0.5<z<3$ \citep{noeske2007,
  elbaz2007, daddi2007, rodighiero2011}. Figure~\ref{fig:ms} 
shows SFR versus stellar mass for our sample. As shown, almost all of
the most massive
ULIRGs follow the MS, implying that these ULIRGs share
similar stellar population properties, while the five least-massive
ULIRGs lie above the MS\null. 
The four X-ray sources lie at the MS (see Figure~\ref{fig:ms}).

\subsection{Morphologies of ULIRGs}\label{sec:morph}

Morphologies of ULIRGs in this sample provide direct information on
how these objects formed and how their intensive star formation was
triggered. It is very challenging to study morphologies of dusty
galaxies at high redshifts. Observed visible light probes the rest-frame UV
bands for objects at $z\sim2$, and therefore their apparent morphologies can
easily be changed by patchy dust extinction. For example, a disk
galaxy at $z\sim2$ with a patchy dust distribution may look like an
irregular galaxy in the visible bands. Recent deep {\it HST} visible
imaging shows that most distant galaxies have apparent irregular
morphologies \citep{abraham1996, abraham2001, lotz2006}, and 
\citet{huang2009} showed that IRAC-selected ULIRGs have irregular,  clumpy
morphologies in the {\it HST}/ACS F814W band.

\begin{figure*}
\centering
\includegraphics[angle=0,width=0.9\textwidth]{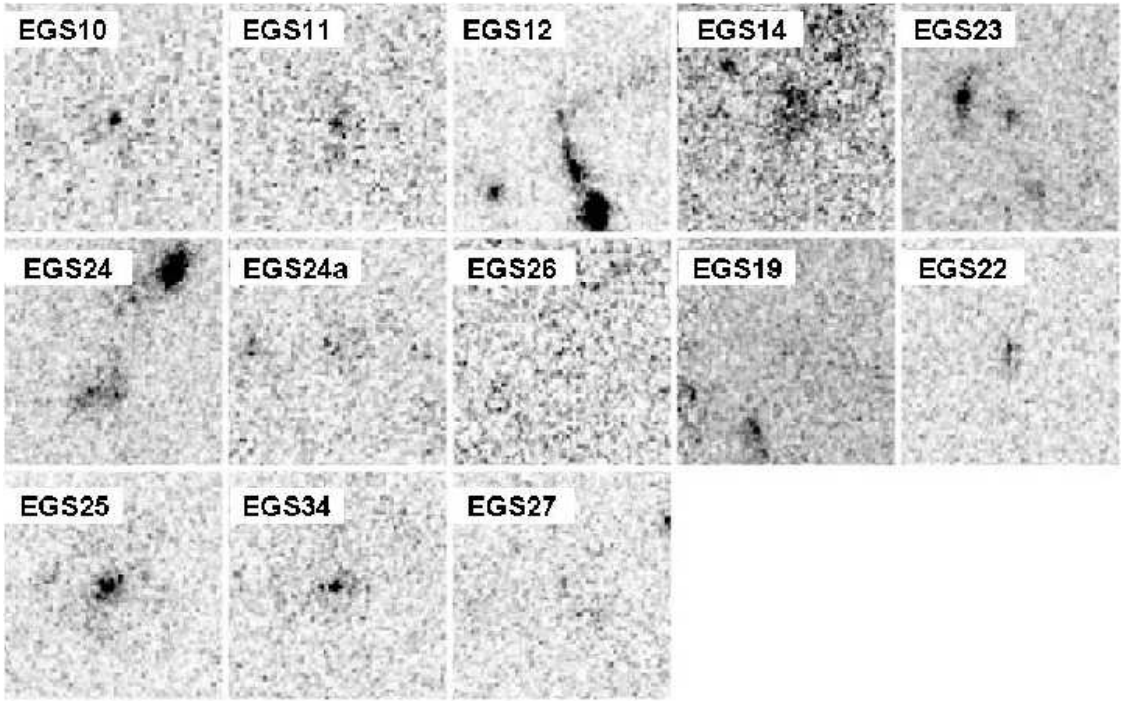}
\caption{ {\it HST}/ACS $I$-band images of 13 ULIRGs. All images are
  in negative grey scale and are 3\arcsec\ square. Source nicknames
  are indicated in each panel.  Five sources are from the present
  sample (Table~1), and the remainder are from \citet{huang2009}.
  \label{fig:mor-I}}
\end{figure*}

\begin{figure*}
\centering
\includegraphics[angle=0,width=0.9\textwidth]{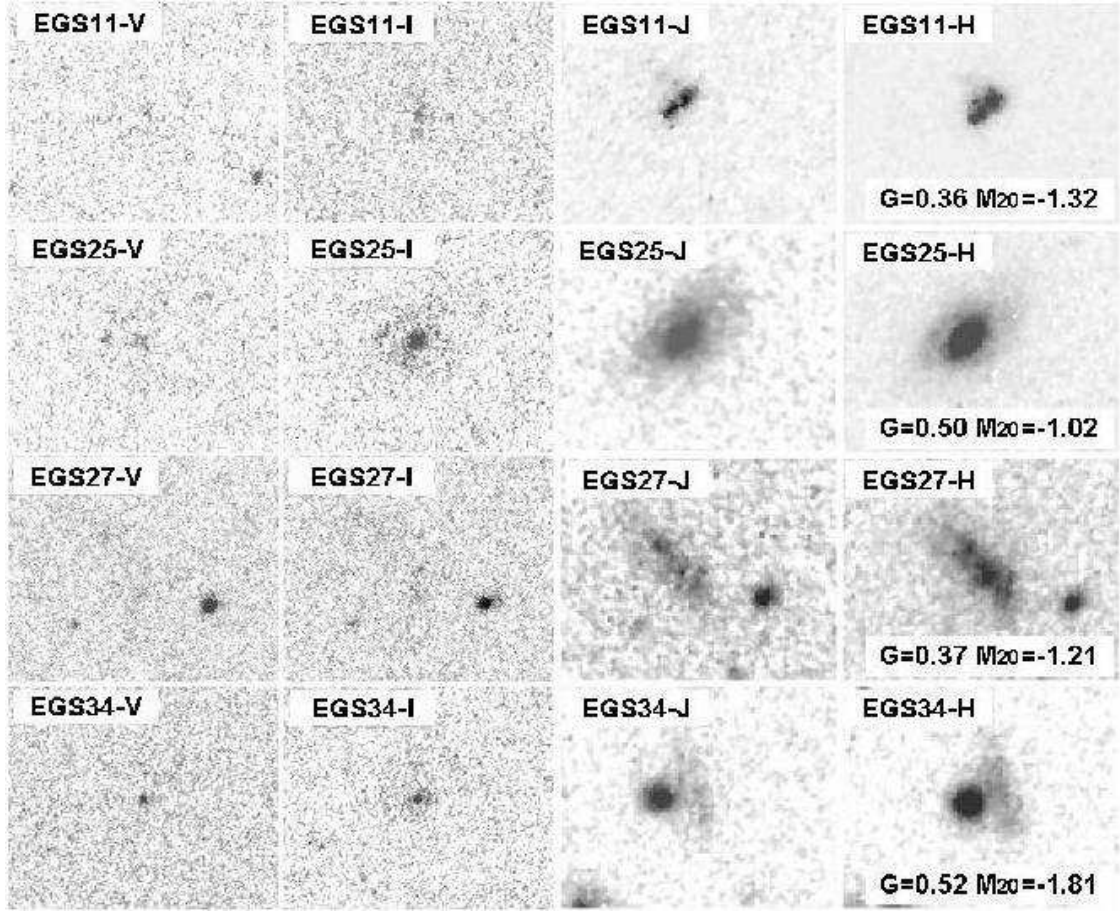}
\caption{ {\it HST}/ACS and WFC3 $V-$, $I-$, $J-$, and $H-$band
images of four ULIRGs. All images are in negative grey scale and are
6\arcsec\ square.  Nicknames are indicated in each panel, and values
of the Gini coefficient $G$ and the $M_{\rm 20}$
are shown in the $H$-band panels.  
The four objects shown are the only ones in the combined
\citet{huang2009} and Table~1 samples that have both ACS and WFC3
imaging.
  \label{fig:mor-H}}
\end{figure*}

The {\it HST}/ACS visible imaging covers the central half of the EGS,
and only a fraction of our IRAC-selected ULIRGs are in the ACS
imaging area. Figure~\ref{fig:mor-I} shows ACS $I$-band
(F814W) stamp images for 14 ULIRGs, five from the present sample and
nine from \citet{huang2007}. All of them
show either extended, irregular morphologies or no detection
However, these rest-UV images view only the hottest stars and can be
heavily affected by dust extinction, and it is therefore essential
to study morphologies in the rest-frame visible, which
shifts to the observed NIR.   CANDELS is  the
largest {\it HST} F125W and F160W imaging survey with the newly installed
NIR camera WFC3.  Its high angular resolution  permits
studying galaxy morphologies even at high redshifts \citep{grogin2011,
  koekemoer2011}. The EGS is one of five fields in
CANDELS\null. Only a part of CANDELS EGS imaging is available
now, and so far only four objects in our sample (EGS11/EGS25/EGS27/EGS34)
are detected at F125W and F160W.    Figure~\ref{fig:mor-H} compares
their visible and NIR morphologies. All four objects are very
red and barely detected in the F814W band, but their NIR
morphologies differ from each other. To describe 
clearly the morphologies of these sources, we have performed 
nonparametric measures of galaxy morphology in the $H$-band images, 
such as Gini coefficient 
(the relative distribution of the galaxy pixel flux values, or $G$) 
and $M_{\rm 20}$ (the second-order moment of the brightest 20\% of the 
galaxy's flux) \citep{lotz2006}. As shown in the $H-$band panels of Figure~\ref{fig:mor-H},
our results are consistent with the $G = 0.4M_{\rm 20} + 0.9$ 
relation defined by \citet{bussmann2011}: Galaxies with $G < 0.4M_{20} + 0.9$ have 
diffuse structures or multiple bright nuclei in appearance (EGS11 and EGS27).
Objects with $G > 0.4M_{20} + 0.9$, they are relatively smooth with 
a single nucleus (EGS25 and EGS34).
\begin{flushleft}
{\bf EGS11} is an IRAC-selected source from \citet{huang2009}. It
is detected in both ACS F606W and F814W though faintly, and its
visible  morphology is hard to discern but consistent with being
point-like.  The NIR morphology in contrast shows a clumpy,
irregular pattern in an overall linear structure. This type of morphology was first discovered in
early {\it HST}  imaging of distant galaxies and thought to
indicate galaxies undergoing intensive star formation \citep{cowie1994}.\\

{\bf EGS25} shows a disk or possibly spiral morphology with a prominent central
bulge. This source is an X-ray sources and identified by its MIPS
$24~\mu$m and 1.4~GHz radio emission  to harbor an
AGN (discussed in Sec.~\ref{sec:agn}). This source is a ULIRG with
AGN yet has an early-type spiral morphology. Usually such a disk galaxy cannot survive
a major merger \citep{DiMatteo2005}. On the other hand, this
galaxy has a SFR as high as 640~\Msun~yr$^{-1}$ compared to
just a few tens of \Msun~yr$^{-1}$ at most in the disks of local
spiral galaxy \citep{kennicutt1998}.  We speculate that the
circumnuclear region is forming stars  through fast
collapse \citep{granato2004, lapi2011} while keeping or
growing the disk at the same time (perhaps through accretion of
high-angular-momentum gas).\\ 

{\bf EGS27} is not detected at F606W and F814W, qualifying it as a
DOG \citep{dey2008}. Its NIR morphology is very extended and
clumpy.  This source is also an X-ray source, but it has no
point-like structure detected at either F125W or F160W.  This 
suggests obscuration, but the MIR spectrum (Fig.~\ref{fig:irs}) does
not exhibit especially strong silicate absorption.\\

{\bf EGS34} shows two distinctively different components within
1\farcs0: one point component and one extended,
low-surface-brightness component. This source is also an X-ray
source. The  point source is detected at ACS F606W and F814W, but the
extended component is very red and is not detected at these
wavelengths. The X-ray
emission is likely from the point source.  The two components could be a
projection  of two objects at different redshifts with the extended  source
contributing little or no MIPS $24~\mu$m emission. 
\end{flushleft}

\subsection{AGN in ULIRGs}\label{sec:agn}

\begin{figure*} \centering
\includegraphics[angle=0,width=0.9\textwidth]{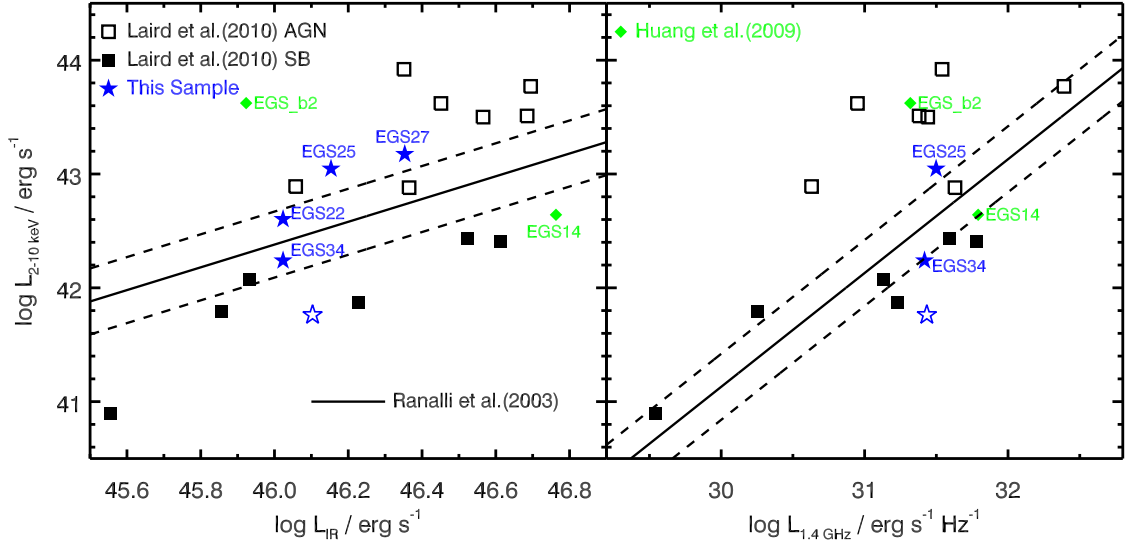}
\caption{ Rest-frame 2--10 keV luminosity for the X-ray-detected
galaxies as functions of SFR indicators.  The left panel shows
$\LIR$, and the right shows rest-frame 1.4 \~GHz radio luminosity.
Labeled stars denote individual sources from our ULIRG sample, and
open stars show the stacking results for the undetected
ULIRGs. EGS22/EGS27 are not in the 1.4~GHz radio catalog. Squares
represent the X-ray-detected SMGs from \citet{laird2010} with filled
squares representing SMGs dominated by star formation and open
squares representing AGN SMGs. Solid lines show the mean local
relations for purely star forming galaxies \citep{ranalli2003},
and dashed lines show a factor of two above and below the mean relations.
\label{fig:X-IR}}
\end{figure*}

One of the surest ways of identifying an AGN is to measure its X-ray
luminosity. Four objects in our sample, EGS22/EGS25/EGS27/EGS34, are
X-ray sources in the Chandra 800~ks AEGIS-X catalog
\citep{nandra2007, laird2009}. Only one (EGS22) of the four
X-ray-selected ULIRGs has $[3.6]-[4.5]>0$ and $[5.8]-[8.0]<0$. Their
X-ray fluxes $F({\rm 0.5-10~keV})$ are $7.7\times10^{-16}$,
$1.7\times10^{-15}$, $1.7\times10^{-15}$, and
$5.7\times10^{-16}$~erg~cm$^{-2}$~s$^{-1}$, respectively,
corresponding to X-ray luminosities $L_{\rm X}$ of
$1.7\times10^{43}$, $2.5\times10^{43}$, $5.6\times10^{43}$, and
$1.0\times10^{43}$~erg~s$^{-1}$. Intensive star formation in ULIRGs,
however, can also generate such a high X-ray luminosity
\citep{laird2010}.  Figure~\ref{fig:X-IR} compares X-ray luminosity
and inferred \LIR. EGS22 and EGS34 have $L({\rm 2-10~keV})/\LIR$
ratios consistent with the $L({\rm 2-10~keV})$--SFR conversion ratio
proposed by \citet{ranalli2003}, while $L({\rm 2-10~keV})/\LIR$
for EGS25 and EGS27 are higher and indicate an AGN
contribution. Only EGS27 has a low X-ray hardness ratio of
$-0.29$. Thus it is a Type~1 (unobscured) AGN based on its X-ray
luminosity and hardness ratio \citep{messias2010}. The remaining 10
ULIRGs in the sample are not detected in X-rays with upper limit
$L_{X}<1.3\times10^{42}$~erg~s$^{-1}$ and thus show no indication
of an AGN.  However, an AGN could still be present if it is
weak or the X-rays are obscured.

\begin{figure}
\centering
\includegraphics[angle=0,width=\columnwidth]{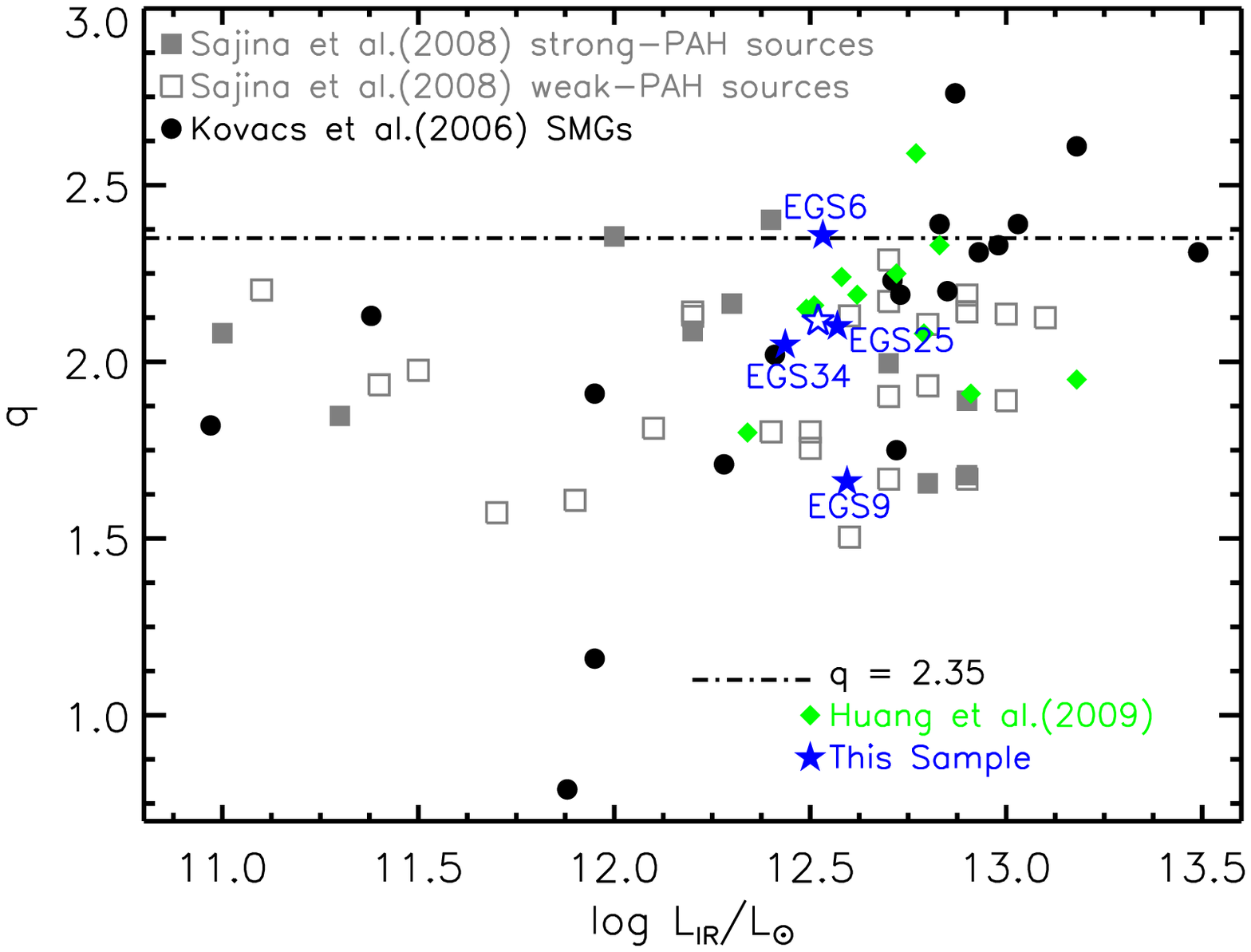}
\caption{Logarithmic FIR to radio ratio as a function of \LIR. The
definition of $q$ is given in Section~\ref{sec:agn}.  Labeled stars
represent individual sources from our sample of ULIRGs (Table~1), and
the open star shows the stacking result for the undetected
ULIRGs. Other symbols show sources from three $z\sim2$ samples
\citep{kovacs2006, sajina2008, huang2009} plotted for comparison. The
dot-dash line indicates $q=2.35$ \citep{yun2001}, typical of starburst-dominated
ULIRGs.
  \label{fig:LIR-q}}
\end{figure}

Radio 1.4~GHz emission can also be used to identify AGNs.
Starburst-dominated IR-luminous galaxies have a typical
$\LIR:L_{\rm 1.4~GHz}$ ratios characterized by $q=2.35$\footnote{ $q\equiv{\rm
    log}({F_{\rm FIR}}/{\rm 3.75\times10^{12}~W~m^{-2}})-{\rm log}({F_{\rm
      1.4~GHz}}/{\rm W~m^{-2}~Hz^{-1}})$ defined by \citet{condon1992}.} \citep{yun2001},
while AGNs have much lower $q$ because the active nuclei emit
non-thermal synchrotron radio radiation but relatively little FIR
radiation. The bright IRAC-selected galaxies in the EGS
\citep{huang2009} have $q\sim2.15$, slightly lower than $q=2.35$ (see
Figure~\ref{fig:LIR-q}). Power-law ULIRGs have much lower $q$ in the
range of $1.6<q<2.15$ \citep{sajina2008, huang2009}, indicating a
much higher AGN fraction. \citet{kovacs2006} measured \LIR\ using
350~\micron, 850~\micron, and 1.2~mm flux densities and obtained a
mean $q=2.07\pm0.3$ for SMGs at $1<z<3$. Four galaxies in our sample,
EGS6/EGS9/EGS25/EGS34, are detected in the 1.4~GHz radio catalog
\citep{ivison2007}.  They have $q=2.36$, 1.66, 2.10, and 2.05,
respectively. Thus only EGS9 shows a strong radio excess.  Even
EGS25, an X-ray AGN, has relatively large $q=2.10$. The remaining 10
ULIRGs in the sample were not detected at 1.4~GHz with radio upper
limits corresponding to $q\ga 2.12$.

Hot dust emission in the rest $3~\mum<\lambda<8~\mum$ wavelength
range is another sign of an AGN \citep{carleton1987, shi2005,
  shi2007}.  The IRAC 8~\micron\ and AKARI 15~\micron\ photometric
data correspond to rest-frame 2.7 and 4.5~\micron, respectively.  The
AKARI 15~\micron\ photometry is rather shallow, and only the three
X-ray-selected sources (EGS25/EGS27/EGS34) were detected.  Their luminosities
$L(4.5~\micron~{\rm rest})$ are $7.8\times10^{10}$, $4.9\times10^{10}$, and
$4.3\times10^{10}~\Lsun$ respectively.  These values are comparable
to the X-ray luminosities with $L_{X}/L(4.5~\mum)= 0.08$, 0.30, and
0.06, consistent with AGN SEDs \citep[e.g.,][]{Elvis1994}. These
galaxies also show red  $[5.8]-[8.0]$
colors, consistent with a steeply-rising AGN continuum. EGS25 has the
highest 15~\mum\ emission in the sample;
its $F({15~\mum})/F({24~\mum})$ ratio is consistent with the ratio for a
QSO at the same redshift \citep{huang2009}.

\begin{figure}
\centering
\includegraphics[angle=0,width=\columnwidth]{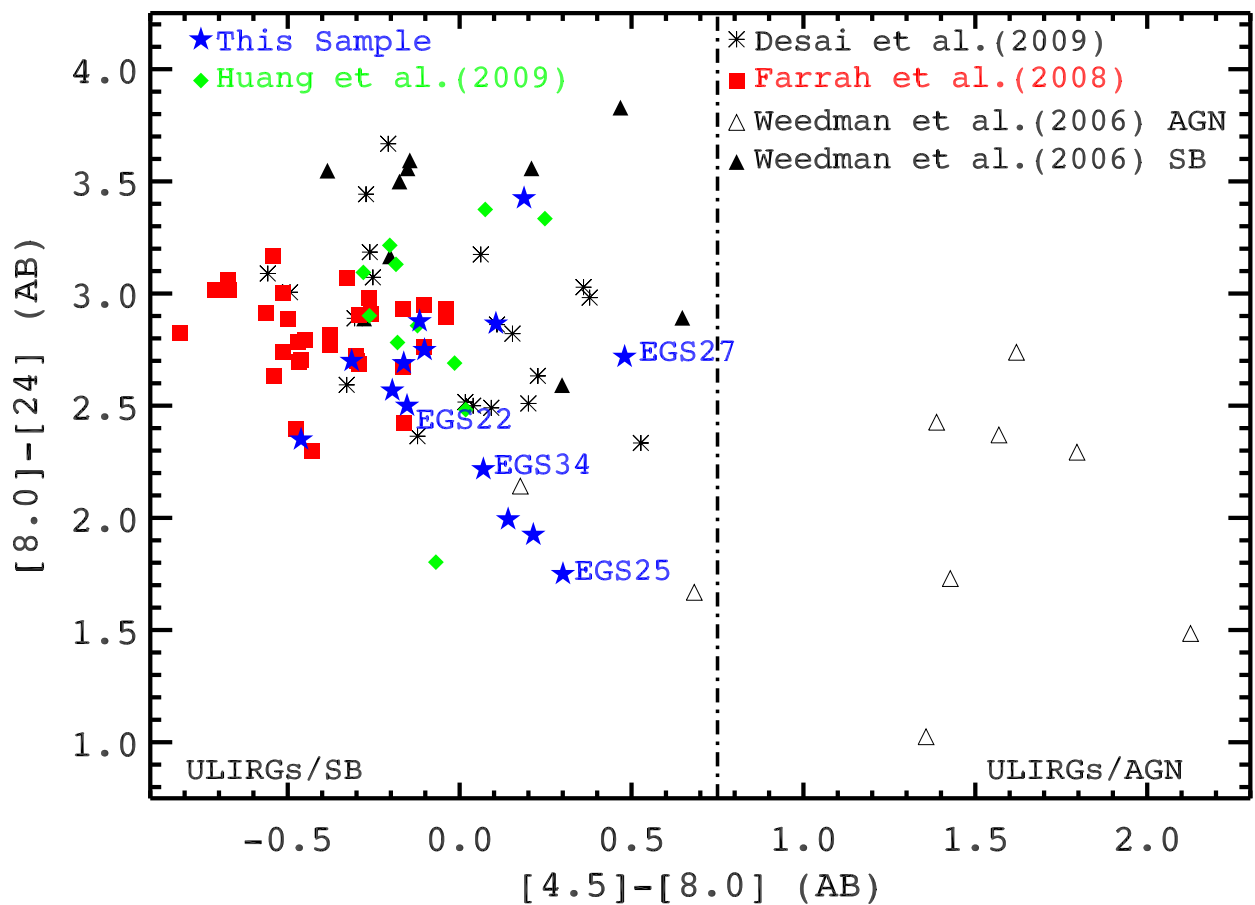}
\caption{ \s~color-color diagram.  Points denote ULIRGs from various
$z\sim2$ spectroscopic samples as indicated in the figure legend.
The four X-ray-detected objects in the present sample are
labeled. The dot-dashed line ($[4.5]-[8.0]=0.75$) was adopted from
\citet{pope2008}.
  \label{fig:AGN-SB}}
\end{figure}

The observed $[4.5]-[8.0]$ color characterizes the ratio of stellar
emission (rest 1.5~\micron\ at $z=2$) to hot dust
emission. Figure~\ref{fig:AGN-SB} shows that seven of nine ULIRGs
\citet{weedman2006} identified as powered primarily by AGNs have
$[4.5]-[8.0]\ga0.75$, the criterion proposed by \citet{pope2008} to
separate AGN- from starburst-dominated SMGs. The \citet{weedman2006}
AGNs also have $[8.0]-[24]\la 2.5$.  In contrast, all the
IRAC-selected ULIRGs \citep{weedman2006, farrah2008, huang2009,
  desai2009} have $[4.5]-[8.0]<0.75$, and most have
$[8.0]-[24]>2.5$. All objects in our sample except the ones detected
in X-rays have these colors, but those detected in X-rays have
$[8.0]-[24]\la 2.5$ colors but still $[4.5]-[8.0]<0.75$.  These
colors may indicate a mix of processes in which objects have both
intensive star formation and AGNs.

\section{Summary}\label{sec:sum}

Applying two IRAC color criteria (inequality 1) to objects 
with $0.2<\FMI<0.6$~mJy produces a sample of ULIRGs that are star-forming, not
AGN-dominated.  The 14 sources  
fall into a narrow redshift range around
$z\sim1.95\pm0.19$ and have PAH features at 7.7, 8.6, and
11.3~\mum. The redshift distribution in our sample is very similar to
that of all SB-dominated ULIRGs ($z\sim2$) selected at 24~\micron.
The objects have stellar masses of $M_*
> 10^{11}~\Msun$, similar to stellar masses of DRGs, BzKs, and
SMGs. Indicated SFRs, based on rest-frame 8~$\mu$m luminosities and
the \citet{caputi2007} $L(8~\mum)$--$\LIR$ relation, are in
the range of 410 to 1100~\Msun~yr$^{-1}$.

Four objects in the sample  are detected in X-rays (three having been
initially selected as X-ray sources), but the X-ray
luminosities for two of them can be accounted for by their
intensive star formation.  The remaining two objects have higher $L_{\rm
  2-10~keV}$ than their star formation can produce, indicating
that they harbor AGNs. MIR colors of all 14 sources in the sample are
consistent with starbursts, but three of the X-ray sources  have
high 4.5~\micron\  luminosities, relatively red
$[4.5]-[8.0]$ colors, and relatively blue $[8.0]-[24]$ colors,
consistent with an AGN contribution to the emission.
FIR/radio ratios for this sample are consistent with $q=2.35$ without
showing strong radio excess from AGN except for one source (EGS9, not one
of the X-ray sources) which has
$q=1.66$. Ten of 14 objects in the sample show no indication that an
AGN is present, and in no object does an AGN appear to dominate the luminosity.

Only four objects in the sample have so far been studied with {\it
  HST}/WFC3 F160W imaging. Their rest-frame visible morphologies are
diverse with one apparent early-type spiral and the others irregular
but of differing descriptions. This diversity suggests that there may
be multiple formation process for ULIRGs, but a larger sample of
imaging is needed (and is now in progress in the CANDELS program) to
reach any strong conclusions.  The observed visible {\it HST}/ACS
imaging probes the rest-frame UV for this sample.  Some objects in
the sample are so red that they are barely detected at F814W\null. Those
that are seen show irregular and clumpy morphologies, consistent with
star formation heavily obscured by patchy dust.

\acknowledgments 

We are grateful to the referee for the comments
which have improved this paper greatly. We thank Robert G.  Abraham,
Christopher J. Conselice and Yu Dai for their valuable
suggestions. This work is based on observations taken by the CANDELS
Multi-Cycle Treasury Program with the NASA/ESA {\it HST}, which is
operated by the Association of Universities for Research in
Astronomy, Inc; under NASA contract NAS5-26555. This work is based
in part on observations made with the {\it Spitzer Space Telescope},
which is operated by the Jet Propulsion Laboratory, California
Institute of Technology under a contract with NASA\null. Support for this
work was provided by NASA through an award issued by JPL/Caltech.
This work was also 
supported by the National Natural Science Foundation of China
(NSFC, No.\ 11225315), the Chinese Universities Scientific Fund (CUSF), and
the Specialized Research Fund for the Doctoral Program of Higher Education
(SRFDP, No.\ 20123402110037).



\begin{table}
\scriptsize
\rotate
\caption{IRS Observation Sample}
\label{tab1}
\vspace{0.1cm}
\begin{tabular}{@{}lcccccccc}
\tableline
\tableline
Nickname\tablenotemark{a}
& EGSIRAC\tablenotemark{b}
& R.A.\tablenotemark{c}
& Dec.\tablenotemark{c}
& $F(3.6~\mum)$
& $F(4.5~\mum)$
& $F(5.8~\mum)$
& $F(8.0~\mum)$
& $F(24~\mum)$\\
&& J2000&J2000& $\mu$Jy & $\mu$Jy & $\mu$Jy & $\mu$Jy & $\mu$Jy  \\
\hline
EGS2  & J142326.59+533457.9&14:23:26.60 & +53:34:58.1 & 27.14$\pm$4.66 &  31.68$\pm$0.64 &  39.19$\pm$6.50 &  27.28$\pm$2.05 &  324.93$\pm$30.41\\
EGS3  & J142255.06+533453.8&14:22:55.07 & +53:34:53.7 & 27.69$\pm$2.86 &  33.26$\pm$0.45 &  36.73$\pm$2.15 &  29.88$\pm$1.73 &  422.85$\pm$10.05\\
EGS5  & J142220.76+532920.1&14:22:20.80 & +53:29:20.2 & 43.02$\pm$9.96 &  50.05$\pm$1.62 &  55.44$\pm$1.83 &  37.48$\pm$1.82 &  450.59$\pm$46.80\\
EGS6  & J142327.26+532819.0&14:23:27.27 & +53:28:19.4 & 65.34$\pm$1.41 &  67.68$\pm$0.99 &  78.46$\pm$2.12 &  82.45$\pm$2.08 &  485.50$\pm$12.20\\
EGS9  & J142246.97+532025.9&14:22:46.96 & +53:20:26.1 & 75.22$\pm$1.72 &102.62$\pm$1.61\0 & 90.71$\pm$2.13 &  67.10$\pm$1.58 &  584.23$\pm$16.45\\
EGS19   & J142118.10+531746.4&14:21:18.12 & +53:17:46.3 & 35.78$\pm$1.06 &  46.31$\pm$0.63 &  53.79$\pm$1.74 &  38.67$\pm$1.32 &  411.70$\pm$35.83\\
EGS21   & J141955.30+530323.2&14:19:55.30 & +53:03:23.2 & 7.59 $\pm$0.72 &  15.14$\pm$0.54 &  20.85$\pm$0.91 &  17.99$\pm$1.20 &  421.68$\pm$36.93\\
EGS22   & J142038.49+525749.8&14:20:38.49 & +52:57:50.0 & 32.66$\pm$4.76 &  42.01$\pm$0.69 &  46.42$\pm$1.00 &  36.49$\pm$2.52 &  364.91$\pm$52.45\\
EGS25   & J141947.52+525026.2&14:19:47.51 & +52:50:26.2 & 62.22$\pm$0.48 &  68.35$\pm$0.37 &  59.24$\pm$1.13 &  90.15$\pm$1.13 &  451.84$\pm$16.20\\
EGS27   & J141935.69+525109.0&14:19:35.71 & +52:51:09.0 & 20.12$\pm$0.99 &  25.50$\pm$0.95 &  35.02$\pm$1.34 &  39.67$\pm$1.31 &  485.05$\pm$12.93\\
EGS32   & J141835.11+524933.9&14:18:35.12 & +52:49:33.7 & 21.68$\pm$1.39 &  29.83$\pm$0.47 &  38.10$\pm$1.50 &  33.97$\pm$1.35 &  213.10$\pm$15.26\\
EGS34   & J141833.21+524241.8&14:18:33.19 & +52:42:42.0 & 36.69$\pm$1.68 &  49.81$\pm$0.78 &  50.23$\pm$0.99 &  53.08$\pm$1.50 &  409.06$\pm$40.66\\
EGS48   & J141601.52+521550.0&14:16:01.53 & +52:15:50.0 & 24.22$\pm$2.80 &  33.97$\pm$0.26 &  41.72$\pm$0.97 &  37.44$\pm$1.12 &  524.55$\pm$18.09\\
EGS49   & J141603.66+522122.6&14:16:03.69 & +52:21:22.7 & 33.38$\pm$1.02 &  42.19$\pm$0.61 &  52.75$\pm$1.80 &  38.40$\pm$1.24 &  482.82$\pm$~8.22\\
\tableline
\end{tabular}
\tablenotetext{a}{Nicknames are the target names in the \s\
  archive and are used for convenience in this paper, but they are
  not official names and should not be used as standalone source
  identifications.}
\tablenotetext{b}{Source name from \citet{barmby2008}.}
\tablenotetext{c}{R.A. and Dec. are the commanded telescope pointing
coordinates, which differ by no more than 0\farcs33 from coordinates given
by \citet{barmby2008}. For each observation, the telescope was
pointed to the commanded coordinates by high accuracy peakup on
nearby \TMASS\ catalog stars observed on the blue peakup array.}

\end{table}

\begin{table}
\footnotesize
\caption{IRS Sample Selection Criteria.}
\label{tab2}
\vspace{0.1cm}
\begin{tabular}{@{\extracolsep{\fill}}lcl}
\tableline
\tableline
Sample & \FMI   & Color criteria \cr
       & (mJy)  &                \cr
\tableline
\citet{houck2005} &  ${>}0.75$ & $\nu F_{\nu}(24~\mum)/\nu F_{\nu}(I) > 60$\\
\citet{yan2007}   &  ${>}0.90$ & $\nu F_{\nu}(24~\mum)/\nu F_{\nu}(I) > 10$ \& 
                           $\nu F_{\nu}(24~\mum)/\nu F_{\nu}(8~\mum) > 3.16$ \\
\citet{weedman2006}&  ${>}0.90$ & $F(\hbox{X})\tablenotemark{a} \ga 10^{-15}$~erg~cm$^{-2}$~s$^{-1}$\\
\citet{weedman2006}&  ${>}0.90$ & IRAC flux density peak at 
                              either 4.5 or 5.8~\mum~(SB)\\ 
\citet{desai2009} (D09)&  ${>}0.50$ & $R-[24] > 14$ Vega mag and a strong rest-frame \\
 &&                             1.6~\mum\ bump in their IRAC SEDs\\
\citet{farrah2008} (Far08)&  ${>}0.50$ & IRAC flux density peak at 4.5 \mum\\
\citet{fiolet2010} (Fio10)&  ${>}0.50$ & IRAC flux density peak at 5.8 \mum\\
\citet{fadda2010} (Fad10)&  ${>}0.14$ & 24~\mum\ sources fainter than 0.5 mJy\\
\citet{huang2009} (H09)&  ${>}0.60$ & $0.05<[3.6]-[4.5]<0.4$ \&
                              $-0.7<[3.6]-[8.0]<0.5$\\
This paper         &  ${>}0.20$ & $0<[3.6]-[4.5]$ \& $[5.8]-[8.0]<0$\\
\tableline
\end{tabular}
\tablenotetext{a}{{\it Chandra} 0.3--8~keV flux density}
\end{table}

\begin{table}
\footnotesize
\caption{Spectroscopic redshifts and \LIR}
\label{tab3}
\vspace{0.1cm}
\begin{tabular}{cccccccc}
\tableline
\tableline
Nickname & 
Redshift\tablenotemark{a} & 
Redshift\tablenotemark{a} & 
$L({\rm 8~\mum})$\tablenotemark{b} & 
\LIR\tablenotemark{c} &  
\LIR\tablenotemark{d} &  
SFR\tablenotemark{e} &
$F({\rm 0.5-10~keV})$\tablenotemark{f}\\
     &  (SB)        &  (ULIRG)    & (\Lsun)  & (\Lsun)  &  (\Lsun)   & (\Msun/yr)  &~erg~cm$^{-2}$~s$^{-1}$  \cr
\tableline
EGS2 &1.97$\pm$0.08 &1.95$\pm$0.02  &11.41 &12.38$\pm$0.35 &12.05$\pm$0.27  &410$\pm$140  &-  \cr
EGS3 &1.81$\pm$0.02 &1.79$\pm$0.08  &11.47 &12.44$\pm$0.08 &12.10$\pm$0.06  &480$\pm$90\0  &-  \cr
EGS5 &1.95$\pm$0.03 &1.92$\pm$0.06  &11.55 &12.52$\pm$0.24 &12.16$\pm$0.19  &570$\pm$190  &-  \cr
EGS6 &1.91$\pm$0.01 &1.90$\pm$0.02  &11.56 &12.53$\pm$0.06 &12.17$\pm$0.05  &590$\pm$90\0  &-  \cr  
EGS9 &1.79$\pm$0.02 &1.77$\pm$0.05  &11.62 &12.59$\pm$0.08 &12.22$\pm$0.07  &680$\pm$130  &-  \cr 
EGS19  &2.02$\pm$0.03 &2.01$\pm$0.08  &11.54 &12.52$\pm$0.21 &12.16$\pm$0.17  &570$\pm$190  &- \cr 
EGS21  &2.01$\pm$0.04 &1.94$\pm$0.10  &11.55 &12.52$\pm$0.24 &12.16$\pm$0.19  &570$\pm$190  &- \cr
EGS22  &1.98$\pm$0.05 &1.94$\pm$0.10  &11.47 &12.44$\pm$0.35 &12.10$\pm$0.28  &470$\pm$160  &$7.7\times10^{-16}$ \cr
EGS25  &1.65$\pm$0.03 &1.63$\pm$0.05  &11.59 &12.57$\pm$0.08 &12.20$\pm$0.06  &640$\pm$110  &$1.7\times10^{-15}$ \cr
EGS27  &2.31$\pm$0.05 &2.29$\pm$0.09  &11.79 &12.77$\pm$0.16 &12.36$\pm$0.12  &\01020$\pm$370  &$1.7\times10^{-15}$ \cr
EGS32  &2.36$\pm$0.05 &2.34$\pm$0.12  &11.44 &12.41$\pm$0.24 &12.07$\pm$0.18  &440$\pm$150  &- \cr
EGS34  &1.76$\pm$0.02 &1.48$\pm$0.11  &11.47 &12.44$\pm$0.17 &12.10$\pm$0.13  &470$\pm$180  &$5.7\times10^{-16}$ \cr
EGS48  &1.89$\pm$0.03 &1.90$\pm$0.06  &11.58 &12.56$\pm$0.13 &12.19$\pm$0.10  &630$\pm$180  &- \cr
EGS49  &1.91$\pm$0.04 &1.90$\pm$0.07  &11.56 &12.53$\pm$0.13 &12.17$\pm$0.10  &580$\pm$170  &- \cr
\tableline
\end{tabular}
\tablenotetext{a}{Redshifts obtained with a starburst (M82) or
  ULIRG (Arp~220) template as indicated.}
\tablenotetext{b}{The rest-frame luminosity $L({\rm 8~\mum})$.}
\tablenotetext{c}{\LIR\ obtained from $L({\rm 8~\mum})$  with the
  \citet{caputi2007} relation.} 
\tablenotetext{d}{$\LIR$ obtained from $L({\rm 8~\mum})$  with an empirical
relation from \citet{bavouzet2008} equation~7.}
\tablenotetext{e}{SFRs from the \citet{caputi2007} relation and  the calibration by Kennicutt (1998):
${\rm SFR}\,(M_\odot\ {\rm yr}^{-1})=4.5\times10^{-44}
\LIR ({\rm erg}\, {\rm s}^{-1})$.}
\tablenotetext{f}{Four objects in our sample, EGS22/EGS25/EGS27/EGS34, are
X-ray sources in the Chandra 800~ks AEGIS-X catalog.}
\end{table}

\end{document}